\documentstyle[preprint,prd,aps]{revtex}
\begin{document}
\draft
\title{Atomic mass dependence of $\Xi^{-}$ and $\overline{\Xi}^{+}$ 
production \\
in central 250~GeV $\pi^-$-nucleon interactions}
\author{
G.~A.~Alves,$^{1}$ S.~Amato,$^{1,}$\cite{by1} J.~C.~Anjos,$^{1}$ 
J.~A.~Appel,$^{2}$ 
J.~Astorga,$^{5}$  T.~Bernard,$^{5,}$\cite{by1a}
S.~B.~Bracker,$^{4}$ L.~M.~Cremaldi,$^{3}$ 
W.~D.~Dagenhart,$^{5}$ C.~L.~Darling,$^{8,}$\cite{by2} R.~L.~Dixon,$^{2}$ 
D.~Errede,$^{7,}$\cite{by3} 
H.~C.~Fenker,$^{2}$ C.~Gay,$^{4,}$\cite{by3a} D.~R.~Green,$^{2}$ 
R.~Jedicke,$^{4,}$\cite{by4} 
P.~E.~Karchin,$^{6}$ C.~Kennedy,$^{8}$ S.~Kwan,$^{2}$ 
L.~H.~Lueking,$^{2}$ 
J.~R.~T.~de~Mello~Neto,$^{1,}$\cite{by5}
J.~Metheny,$^{5}$ R.~H.~Milburn,$^{5}$ 
J.~M.~de~Miranda,$^{1}$ H.~da~Motta~Filho,$^{1}$ A.~Napier,$^{5}$ 
D.~Passmore,$^{5}$ A.~Rafatian,$^{3}$ A.~C.~dos~Reis,$^{1}$ 
W.~R.~Ross,$^{8,}$\cite{by6} 
A.~F.~S.~Santoro,$^{1}$ M.~Sheaff,$^{7}$ M.~H.~G.~Souza,$^{1}$ 
C.~Stoughton,$^{2}$ M.~E.~Streetman,$^{2}$ 
D.~J.~Summers,$^{3}$ S.~F.~Takach,$^{6}$ 
A.~Wallace,$^{8}$ and 
Z.~Wu$^{8}$ 
\\
(Fermilab E769 Collaboration) \\}
\address{
$^{1}${\it Centro Brasileiro de Pesquisas F\'\i sicas, Rio de Janeiro,
Brazil}
\\
$^{2}${\it Fermi National Accelerator Laboratory, Batavia, Illinois 60510} 
\\ 
$^{3}${\it University of Mississippi, University, Mississippi 38677} 
\\ 
$^{4}${\it University of Toronto, Toronto, Ontario, Canada M5S 1A7}  
\\                                           
$^{5}${\it Tufts University, Medford, Massachusetts 02155} 
\\ 
$^{6}${\it Wayne State University, Detroit, Michigan 48202} 
\\ 
$^{7}${\it University of Wisconsin, Madison, Wisconsin 53706} 
\\ 
$^{8}${\it Yale University, New Haven, Connecticut 06511}}

\date{\today}
\maketitle
\begin{abstract}
We present the first measurement of the atomic 
mass dependence of 
central $\Xi^{-}$ and $\overline{\Xi}^{+}$ production.  
It is measured using a sample of 
22,459 $\Xi^{-}$'s and $\overline{\Xi}^{+}$'s
produced in collisions between a 250~GeV $\pi^-$ beam and
targets of beryllium, aluminum, copper, 
and tungsten.  The relative cross sections
are fit to the two parameter function $\sigma_0 A^\alpha$, where 
$A$ is the
atomic mass.  We measure $\alpha = 0.924\pm0.020\pm0.025$, for 
Feynman-$x$ in the range $-0.09 < x_F < 0.15$. 
\end{abstract}


\narrowtext

% body of paper here

The atomic mass dependence of strong interaction cross sections
with nuclear targets is sensitive to the behavior of hadrons
and quarks inside nuclear matter.  In addition, knowledge of
this dependence is needed to compare cross section results
from experiments using
different target materials.  Many atomic mass dependence
measurements have been made 
\cite{review}.  Nevertheless, little exists in the literature
for central hyperon production.
We report here, the first measurement
of the atomic mass
dependence of central $\Xi^{-}$ and $\overline{\Xi}^{+}$ production.

The atomic mass dependence of cross sections is 
frequently parameterized as
\begin{equation}
\sigma(A) = \sigma_0 A^\alpha,
\end{equation}
where $A$~is the atomic mass of the target.
By using four different target materials, we are able to
check the applicability of this parameterization,
as well as making a measurement
of $\alpha$.

A model of the nucleus as a totally absorbing sphere gives a value
for $\alpha$ of 0.67.  For absorption cross sections, $\alpha$ is
a little higher. For example, Carroll {\it et al.} \cite{one} 
measured $\alpha = 0.755\pm0.010$
for a 280~GeV~$\pi^-$ beam.
If a cross section were simply 
proportional to the number of nucleons in the nucleus, $\alpha$ 
would be
1.00.  Earlier, we reported a measured value for $\alpha$ of 
$1.00\pm 0.05\pm0.02$
for D meson production \cite{two}. 

The apparatus in Fermilab experiment E769 has been previously 
described (see~\cite{three} and references 
therein).
The targets were 26 foils of Be, Al, Cu, and W with a total nuclear
interaction length of 2\%.  The foils were simultaneously exposed to
the 250~GeV beam to minimize errors associated with flux 
measurement.  A differential \v{C}erenkov counter was used to measure
the beam content and reduce contamination from kaons.  The final 
sample consisted of 95\%~$\pi^-$ with a contamination of 
3\%~$K^-$ and 2\%~$\overline{p}$.

The elements of the spectrometer relevant to this analysis are
11 silicon microstrip planes (1--30~cm 
downstream of the targets), 35 drift chamber planes (150--1750~cm), 
2 multiwire proportional chambers (130~cm, 180~cm),
and 2 magnets (290~cm, 620~cm) 
for momentum measurement.  The electromagnetic
and hadronic calorimeters were used only for on-line event selection.
The two threshold \v{C}erenkov counters downstream
of the target were not 
used in this analysis.

Tracks of charged particles were reconstructed using hits in the 
detector planes.  The $\Xi^-$'s were reconstructed using only
the three final tracks produced in the decays 
$\Xi^-\rightarrow\Lambda + \pi^-$ and 
$\Lambda \rightarrow \pi^- + p$ (charge conjugates are
implied in this paragraph and the following paragraph).
The analysis focuses on events where both decays occurred between 
the silicon microstrip planes and the drift chamber planes.  The
tracks used to reconstruct $\Xi^-$'s were required to have hits
only in the drift chambers and proportional chambers, no hits 
being allowed in the silicon microstrip detectors.

We applied further criteria to select $\Xi^-$'s from the 
data.  The reconstructed $\Lambda$ mass was required to be 
within 5.25~MeV of the known mass. The shortest
distance between the two tracks used to 
reconstruct the neutral
$\Lambda$ was required to be less than 0.7~cm.  The 
shortest distance between
the $\Lambda$ track and the other charged pion track
was required to be less than 0.66~cm.
The angle between the $\Xi^-$ trajectory and the direction from the 
primary vertex to the $\Xi^-$ decay vertex was
required to be less than 0.012~radians.  There were
requirements  on the
geometric locations of the three vertices and on the charges
of the three decay tracks.  These were the most significant
selection criteria.

Figure \ref{massplot} shows the invariant mass distribution
for candidate $\Xi^-$'s and $\overline{\Xi}^+$'s after the
selection criteria were applied.  The figure shows a strong 
signal over a linear background.  
The signal was determined using sideband subtraction, not
by fitting a function to the mass distribution.
This eliminates errors associated with the determination of 
the shape of the signal peak.

The combined $\Xi^-$ and $\overline{\Xi}^+$ data signals, before
acceptance corrections and weighting are applied, are as follows: 
$1980\pm55$ from a minimum bias trigger 
(no requirement on 
transverse energy
in the calorimeters), and $20479\pm187$ from a trigger that 
required greater than roughly 5.5~GeV of 
transverse energy in 
the calorimeters.  The data acquisition rates for these triggers
were controlled using prescalers which were set to record
a specific fraction of events that passed trigger requirements.
The signals from the two triggers are combined 
using weights based on the known prescaler settings.
Typically, the minimum bias events with less than 5.5 GeV
of tranverse energy have weights roughly 20 times larger
than the events from the transverse energy trigger.
This causes the statistical
error on $\alpha$ to be dominated by the statistical errors on the
events from the minimum bias trigger.

A full detector simulation was used to calculate the acceptances 
for each
material in narrow bins of $x_F$ (width 0.015).  
The acceptance calculation was repeated for each data set listed 
in Table~\ref{ares7}.
Acceptances ranged from 3\% to 12\%.
The simulation modeled the geometry of
the detector, the primary interaction, secondary interactions, 
pair production, multiple scattering, detector plane 
efficiencies, and all analysis cuts.  A total of 
1.8 million simulated
events were generated for the acceptance calculation.
The statistical errors on the acceptances were much smaller 
than the statistical errors from the data.

The dominant systematic error is due to the uncertainty in
simulating the average number of charged particles per event
(the multiplicity).  The average multiplicity increases with 
the atomic mass of the target.  Since the targets were arranged
along the beam direction in order of decreasing atomic mass,
events produced in the higher mass targets also
suffered more pair
production and secondary interactions than those produced
in low mass targets.  The overall effect is to reduce the
acceptance for the higher mass targets, because the track
reconstruction efficiency is lower in high multiplicity events.
This systematic error was studied by varying the multiplicity
of generated events in the simulation and by making
comparisons between the data and the simulation.
The systematic error on $\alpha$ related to multiplicity
was estimated to be $\pm0.023$.  The systematic
error was somewhat higher 
in events where
the $\Xi^-$'s and $\overline{\Xi}^+$'s had low transverse
momenta ($p_T$), because the track densities were higher 
nearer the beam.

We studied several other sources of systematic error.
Systematic errors in  $\alpha$ associated with measurement 
of the thickness
of the target foils, and errors associated with
the location of the reconstructed primary vertex
were each estimated to be roughly $\pm0.007$.
Other systematic errors were estimated to be even smaller.
These include errors related to inelastic collisions in the target
that attenuate the beam flux, simulated $\Xi^{-}$ and
$\overline{\Xi}^{+}$ momentum distributions, detector geometry,
signal determination, beam contamination, and trigger biases.

The values of $\alpha$ were determined by fitting the two parameter 
function $\sigma_0 A^\alpha$ to four data points: 
the relative Be, Al, Cu, and W cross sections.   The 
fit parameter $\sigma_0$ simply
normalizes the function.  Figure~\ref{crosssect}
shows one of the fits.  The $\chi^2$ of this
fit is 1.35 with two degrees of freedom.  The 
function fits our data well.

Table~\ref{ares7} shows measured values of $\alpha$
for several data sets. 
The largest data set covers the region in $x_F$
where the acceptance is large enough to yield enough 
statistics for a meaningful result.  This large data set 
is subdivided in several different ways.
When calculated separately for $\overline{\Xi}^{+}$ and 
$\Xi^{-}$, $\alpha$ is the same within errors. 
As a function of both $p_T$ and $x_F$, the dependence of
$\alpha$ is consistent with being flat within errors, 
but there is a rise near $x_F = 0.00$ and in the highest
bin of $p_T$.
Note that the systematic errors are strongly 
correlated in different 
subsets of data.  In evaluating trends as a function of $x_F$ or 
$p_T$, the 
statistical errors are more important than the systematic errors.

It is useful to compare our results to 
other experimental measurements.
In \cite{pikp,pikp2}, the atomic mass dependence
for production of $\pi^\pm$, $\mbox{K}^\pm$, p and 
$\overline{p}$ was reported.  That experiment used
a proton beam to study central production
as a function of $p_T$ and beam energy.  For example, at 
$p_T = 0.77$~GeV with a 400~GeV beam,
$\alpha$ was measured to be $0.91\pm0.01$ for production of
$\pi^-$ and
$0.98\pm0.02$ for production of both $\mbox{K}^-$ and p.
Our result for the atomic mass dependence of 
central $\Xi^{-}$ and $\overline{\Xi}^{+}$ production
is similar
in that $\alpha$ is also a little less than 1.00.
In Ref.~\cite{caszero}, the $\Xi^0$ atomic mass dependence was
reported for $0.2 < x_F < 0.8$ using a 400~GeV proton beam.
The $\Xi^0$ is the particle most similar to the $\Xi^-$.
Figures 31 and 33 of Ref.~\cite{caszero} show $\alpha$ as a 
function of $p_T$ and $x_F$.  These figures show the 
long established trends of $\alpha$ increasing as $x_F$ decreases
towards 0.0
and $\alpha$ increasing as $p_T$ increases.  The $x_F$ figure shows
$\alpha$ increasing from less than 0.5 towards a value
a little less than 1.00
as $x_F$ decreases from 0.8 to 0.0 (no data for
$x_F < 0.2$ in~\cite{caszero}).  This is 
consistent with our measured value for $\alpha$.

In summary, we present the first measurement of the atomic 
mass dependence of the cross section for
central $\Xi^{-}$ and $\overline{\Xi}^{+}$ production.
We measure $\alpha = 0.924\pm0.020\pm0.025$, having found
that $A^\alpha$ is a good parameterization of the atomic mass
dependence of the production cross section.  
Our data are consistent with
the value of $\alpha$ approaching 1.0 as $x_F$
approaches 0.0.

We gratefully acknowledge funding from 
the U.S. Department of Energy, 
the U.S. National Science Foundation, 
the U.S. National Science Foundation Graduate Fellowship Program,
the Wayne State University High Energy Physics Initiative,
the Brazilian Conselho Nacional de
Desenvolvimento Cient\'\i fico e Tecnol\'{o}gico, and the National
Research Council of Canada.

% now the references. delete or change fake bibitem. delete next three
%   lines and directly read in your .bbl file if you use bibtex.

% \begin{figure}
% \caption{}
% \label{}
% \end{figure}

%To create this figure run /d5/people/wdd/cascade/data/m72bins.kumac
%This calls plot3.kumac that formats the figure
\begin{figure}
\caption{Invariant mass distribution for  
$\Xi^{-}$ and $\overline{\Xi}^{+}$ candidates.  
Mass is calculated using the known $\Lambda$ mass, the known 
$\pi^-$ mass, and the measured momenta of the 
three tracks (assumed to be $p$$\pi^-$$\pi^-$ or 
$\overline{p}$$\pi^+$$\pi^+$).  
Only candidates that pass the analysis cuts are included.
Only statistical errors are shown.
Events are weighted based on trigger prescalers.
The average weight is 1.56.
}
\label{massplot}
\end{figure}

%To create this figure run /d5/people/wdd/cascade/final/apaper.kumac
%This calls plotpaper.kumac that formats the figure
\begin{figure}
\caption{Relative cross sections for $\overline{\Xi}^{+}$ and 
$\Xi^{-}$ production ($x_F$ between $-0.09$  
and $0.15$, summed over $p_T$) as a function of the 
atomic mass of the target.
The solid curve shows the fit of the function 
$\sigma = \sigma_0 A^\alpha$, where $\sigma_0$ and $\alpha$ are 
the fit parameters.  Only statistical errors are 
shown and used in the fit.
}
\label{crosssect}
\end{figure}

% \begin{table}
% \caption{}
% \label{}
% \end{tabular}
% \end{table}
\begin{table}

\caption{$\alpha$ with statistical 
then systematic errors.
The overall data set contains $\overline{\Xi}^{+}$'s and 
$\Xi^{-}$'s with $x_F$ between $-0.09$  
and $0.15$, summed over $p_T$.  The other
lines show results from
subsets of the overall data set with the 
additional selection specified
in the first column.}

\label{ares7}

\begin{tabular}{lr@{}l@{${}\pm{}$}r@{}l@{${}\pm{}$}r@{}l}
\multicolumn{1}{c}{Data Set} & 
\multicolumn{6}{c}{$\alpha$ in $\sigma = \sigma_0 A^\alpha$} \\
\hline
 & \\
Overall                                  &  0&.924 & 0&.020 & 0&.025 \\
 & \\
$\overline{\Xi}^+$ only                  &  0&.905 & 0&.028 & 0&.025 \\
$\Xi^-$ only                             &  0&.939 & 0&.026 & 0&.025 \\
 & \\
$p_T$ 0.0 to 0.5~GeV                     &  0&.906 & 0&.041 & 0&.035 \\
$p_T$ 0.5 to 1.0~GeV                     &  0&.913 & 0&.028 & 0&.022 \\
$p_T$ 1.0 to 1.5~GeV                     &  0&.988 & 0&.041 & 0&.020 \\
 & \\
$x_F$ $-0.09$ to $-0.03$                 &  0&.881 & 0&.042 & 0&.025 \\
$x_F$ $-0.03$ to $0.03$                  &  0&.981 & 0&.029 & 0&.025 \\
$x_F$ $0.03$ to $0.09$                   &  0&.910 & 0&.034 & 0&.025 \\
$x_F$ $0.09$ to $0.15$                   &  0&.918 & 0&.056 & 0&.025 \\
  & \\
\end{tabular} 
\end{table}

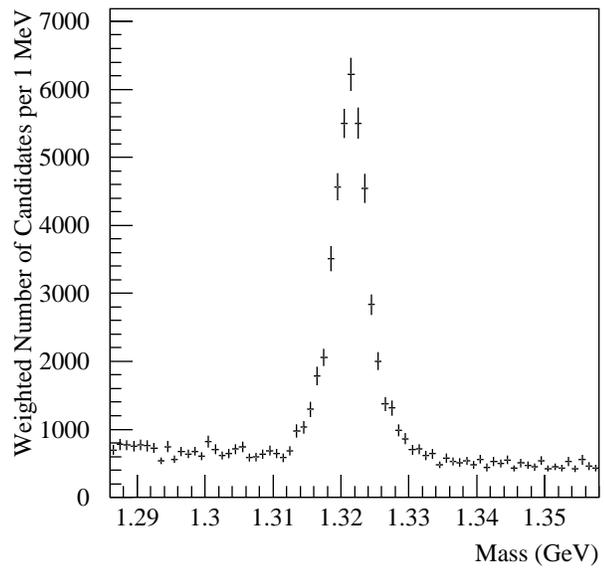
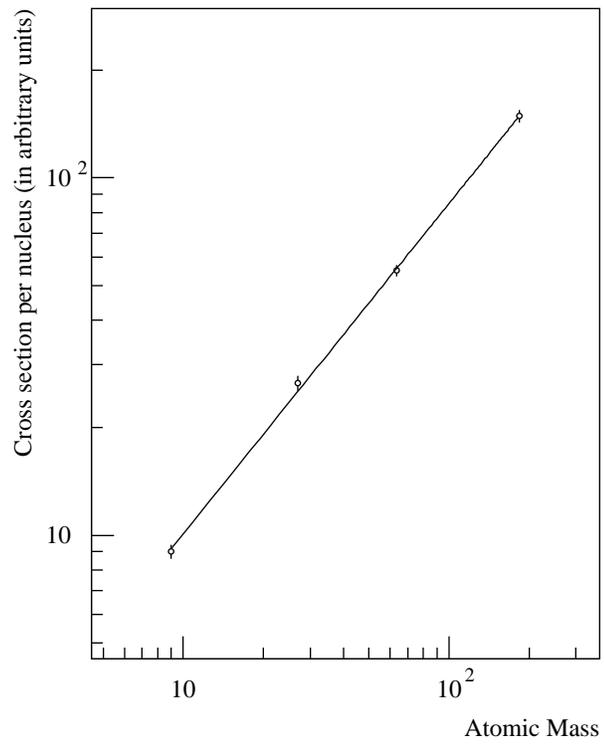
\end{document}